\documentclass[final,leqno,onefignum,onetabnum]{siamltex1213}

\usepackage{amsmath, graphicx, hyperref, amssymb, subcaption, cite}
\usepackage{color}
\usepackage[normalem]{ulem}

\renewcommand{\d}{\partial}
\newcommand{\D}{\mathrm d}
\newcommand{\E}{\mathrm e}

\newcommand{\pD}{\partial }

\title{Operator Splitting Method for Simulation of Dynamic Flows in Natural Gas Pipeline Networks}

\author{Sergey~A.~Dyachenko\footnotemark[2]  \and
Anatoly~Zlotnik\footnotemark[3]  \and
Alexander~O.~Korotkevich\footnotemark[4]  \and
Michael~Chertkov\footnotemark[5] }

\begin{document}
\maketitle

\renewcommand{\thefootnote}{\fnsymbol{footnote}}

\footnotetext[2]{Department of Mathematics, University of Illinois Urbana-Champaign,
1409 W. Green Street, Urbana,  IL 61801, USA; current address: ICERM, Brown University, Box 1995,
Providence, RI 02912, USA.}
\footnotetext[3]{Theoretical Division, T-5,
Los Alamos National Laboratory
Los Alamos, NM 87545, USA.}
\footnotetext[4]{Department of Mathematics \& Statistics, The University of New Mexico,
MSC01 1115, 1 University of New Mexico, Albuquerque, New Mexico, 87131-0001, USA;
L.\,D.~Landau Institute for Theoretical Physics, 2 Kosygin Str., Moscow, 119334, Russia.}
\footnotetext[5]{
Theoretical Division, T-4 \& CNLS,
Los Alamos National Laboratory
Los Alamos, NM 87545, USA;
Energy System Center, Skoltech, Moscow, 143026, Russia.}

\renewcommand{\thefootnote}{\arabic{footnote}}

\begin{abstract}
We develop an operator splitting method to simulate flows of isothermal compressible natural gas over transmission pipelines.
The method solves a system of nonlinear hyperbolic partial differential equations (PDEs) of hydrodynamic type for mass flow and pressure on a metric graph, where turbulent losses of momentum are modeled by phenomenological Darcy-Weisbach friction.
Mass flow balance is maintained through the boundary conditions at the network nodes, where natural gas is injected or withdrawn from the system.
Gas flow through the network is controlled by compressors boosting pressure at the inlet of the adjoint pipe.
Our operator splitting numerical scheme is unconditionally stable and it is second order accurate in space and time.
The scheme is explicit, and it is formulated to work with general networks with loops.
We test the scheme over range of regimes and network configurations, also comparing its performance with 
performance of two other state of the art implicit schemes.
\end{abstract}



\pagestyle{myheadings}
\thispagestyle{plain}
\markboth{S.\,A.~Dyachenko, A.~Zlotnik, A.\,O.~Korotkevich, M.~Chertkov}{Operator Splitting for Pipeline Simulation}

\section{Introduction} \label{sec:intro}

Economic and technological changes have driven an increase in natural gas usage for power generation, which has created new challenges for the operation of gas pipeline networks.
Intermittent and varying gas-fired power plant activity produces fluctuations in withdrawals from natural gas pipelines, which today experience transient effects of unprecedented intensity \cite{CBL2015}.
As a result, there has been renewal of interest in the past decade to methods for accurate simulation of transient flows through pipeline systems \cite{BHK2006,banda08,HMS2010}.

Modified Euler equation and continuity equation describe flow of isothermal compressible gas through a system/network of one-dimensional pipes.
The Euler equation is modified to account for the loss of gas momentum due to turbulence via the phenomenological Darcy-Weisbach formula \cite{wylie78}.
This paper, built of the previous research reviewed in \cite{thorley87,hudson06}, focuses on numerical analysis of the resulting system of PDEs describing unbalanced flow of natural gas over networks that span thousands of kilometers and temporal variations over temporal scales ranging from tens of seconds to hours.

In general, the numerical methods for gas flow simulation are categorized into two classes based on the relation between gas velocity $u$ and the speed of sound $c_s$. The first class of methods solves the full isothermal gas dynamics equations and keep the nonlinear self-advection. The method~\cite{CWShu1998} allows to simulate gas dynamics in the regimes where $u \sim c_s$, and are high-order accurate in smooth regions and essentially non-oscillatory (ENO/WENO) for solution discontinuities. The methods proposed in \cite{ZA2000,BHK2006,herty08} are total variation diminishing (TVD), and are also suitable to simulate shock wave formation and propagation. These highly advanced explicit methods are designed to capture nonlinear transients of shock/emergency type at very short timescales but are impractical and unnecessarily complicated for simulation of gas pipelines in the regimes of normal operations.
%In the regime of interest the friction forces dissipate shock waves and other singularities sufficiently fast. It is not necessary for the discontinuities to be resolved exactly as long as the induced errors do not extend into the regions of smooth solution and are not persistent in time.
%In addition, shockwave phenomena are not typical in the regular pipeline operations, and are indeed avoided, because they would appear immediately after a catastrophic event causing rapid depressurization \cite{marchesin83}.
The second class of methods is designed to work deep inside the normal operation regimes where, $u \ll c_s$. The methods resolves variations in pressure and mass flow on time scales much longer than the rate of acoustic propagation \cite{osiadacz84,kralik88,HMS2010,grundel14}.  The nonlinear advection term in Euler equation, being of the order $u^2/c^2_s$ is omitted and the result is a linear second order PDE (wave equation) with nonlinear damping also known as the Weymouth equation, see e.g. \cite{chua82,Osiadacz1984}.
In this normal operation regime the sound waves (of the preasure/density variations) are largely overdamped and hence the natural choice of numerical methods would be an implicit $L$-stable method. While very efficient to describe slow changes the methods fails to capture wave transients initiated by fast exogenous changes, still typical for normal operation.  (See related discussion in \cite{Kiuchi1994}.)

In this paper we propose new method referred to as ``split-step''. The method is based on operator-splitting technique proposed by G.\,Strang~\cite{strang68}. It has been successfully implemented to simulate the nonlinear Schr\"odinger equation arising, e.\,g., in fiber optics \cite{TA1984,Agrawal2001}. The ``split-step'' method is explicit and unconditionally stable therefore
filling the gap between the two aforementioned classes of methods. Furthermore, the method is readily extendable to complex pipeline networks with nodally-located time-varying compressors that boost gas flow into adjoining pipes. The method is second order accurate in time, and it can be modified to enhance the order of accuracy.  Conservation of total mass of gas in the simulated system is an intrinsic property of the numerical scheme.

The manuscript is organized as follows.
Section \ref{sec:physics} introduces phenomenology, approximations and assumptions underlying the gas dynamics equations.
Then in Section \ref{sec:model} we describe the mathematical setting of the resulting system of  hyperbolic PDEs on a metric graph.
A comprehensive description of the proposed split-step numerical method is provided in Section \ref{sec:method}.
In Section \ref{sec:examples} we present results of numerical simulations using the split-step method in several different settings, and compare the split-step to other methods \cite{Kiuchi1994,ZCB2015}. We provide a summary and discuss directions for future research in Section \ref{sec:conc}.

\section{Gas Dynamics in a Pipe} \label{sec:physics}

Microscopic equations of compressible fluid dynamics consist of the Euler equation and continuity equation. Typically natural gas moves with the average speed which is significantly less than the speed of sound, $c_s$. The gas flow is turbulent (with the Reynolds number typically in the, $\mathcal{R} \sim 10^3\div 10^5$, range). We are interested in modeling transportation of gas over distances which are significantly larger than diameter of the pipe, $D$, we consider averaging over turbulent fluctuations at the scale, $D$, and smaller. The resulting  system of equations describing evolution of  the averaged density, $\rho(t,x)$, and the averaged gas velocity, $u(t,x)$, in time $t$, and space (position along the pipe), $x$, is~\cite{Osiadacz1989}
\begin{align}
& \dfrac{\d \rho}{\d t} + \dfrac{\d (\rho u)}{\d x} = 0, \label{eq:mass1}\\
& \dfrac{\d \rho u}{\d t} + \dfrac{\d (\rho u^2)}{\d x} + \dfrac{\d p}{\d x} =-\dfrac{fc_s^2}{2D}\dfrac{\rho u |\rho u|}{p}, \label{eq:momentum1}
\end{align}
where the averaged pressure, $p(t,x)$, is expressed via $\rho$ according to the standard thermodynamic isothermal, ideal gas relation,
$p = c_s^2 \rho$. Phenomenological Darcy-Weisbach (D--W) term on the right hand side of Eq.~(\ref{eq:momentum1}) models
resistance of the pipe due to the loss of momentum in the result of scattering of sound (variations of density) on turbulent fluctuations.

In the regime of normal operation, excluding very fast transients, the pressure gradient term in Eq.~(\ref{eq:momentum1}) is mainly balanced by the D--W term, while contribution of other terms to the balance is much smaller. In fact, the second self-advection term on the left hand side of Eq.~(\ref{eq:momentum1}) is also smaller than the first term \cite{Osiadacz1984}. Ignoring the second term on the left hand side of the momentum Eq.~(\ref{eq:momentum1}), and expressing $p$ via $\rho$ through ideal gas relation $p=c_s^2\rho$, one arrives at the Weymouth system of equations
\begin{align}
& \dfrac{\d p}{\d t} + c_s^2\dfrac{\d (\rho u)}{\d x} = 0, \label{eq:mass2}\\
& \dfrac{\d (\rho u)}{\d t} + \dfrac{\d p}{\d x} = -\dfrac{fc_s^2}{2D}\dfrac{\rho^2 u|u|}{p},\label{eq:momentum2}
\end{align}
used to model slow transients in a pipe \cite{Kiuchi1994,ZA2000}.
% {\red \sout{They can also be used to describe rapidly changing flows with waves that propagate with frequency greater than $\omega_0$.
% This is a special regime where damping of the acoustic waves is weak, and it is realized when acoustic waves propel turbulent vortices
% or pulsations to dissipation at the boundaries of the pipe.}}{\green We don't want to touch this.}
% Because {\red typical frequency} $\omega_0 \approx 10 \, \mathrm{s}^{-1}$ for  large gas transmission pipelines, the regime of visible acoustic waves
% is realized only in the case of abrupt changes, i.e., depressurization, occurring in tens of a second or at even shorter time scales \cite{marchesin83}.
% This regime of ultra-fast or extreme changes is beyond the scope of our analysis.  %, are discussed in \cite{LCB2016}.
It is convenient, following \cite{Kiuchi1994}, to state \eqref{eq:mass2}-\eqref{eq:momentum2} in terms of the re-scaled time, $t\to t c_s$, and also transition from the velocity variable to the re-scaled mass flow variable, $\phi=c_s\rho u$. The resulting hyperbolic system of equations becomes
\begin{align}
& p_{t} + \phi_x = 0, \label{eq:mass3} \\
& \phi_{t} + p_x = -\dfrac{f}{2D}\dfrac{\phi|\phi|}{p}, \label{eq:momentum3}
\end{align}
where standard shortcut notations for derivatives, $p_x=\d p/\d x$ and $p_t=\d p/\d t$, were used.

\section{Gas Dynamics over Network} \label{sec:model}

The system of \eqref{eq:mass3}-\eqref{eq:momentum3}, governing dynamics of pressure and mass flow in a pipe, constitutes a building block for describing dynamics of pressure and mass flow in a network/graph of pipes, ${\cal G}=({\cal V},{\cal E})$, where ${\cal V}$ and ${\cal E}$ stand for the set of nodes and the set of undirected edges of the graph respectively. Pressures and mass flows at different pipes are linked to each other through a set of boundary conditions. First of all, one requires that at any node of the system the mass is conserved:
\begin{eqnarray}
\forall i\in{\cal V}:\quad \sum_{j:\{i,j\}\in{\cal E}} S_{ij}\phi_{ji}(t,L_{ij})=q_i(t),\label{eq:nodalflowbal0}
\end{eqnarray}
where $\phi_{ij}(x,t)$, with $x\in [0;L_{ij}]$, denotes the mass flow (per unit area) within the pipe $\{i,j\}$ of length, $L_{ij}$, and the cross-section area, $S_{ij}$, with the convention that, $\phi_{ji}(L_{ij},t)=-\phi_{ij}(t,0)$, is positive/negative if the gas leaves/enters the pipe $\{i,j\}$ at the node $i$. $q_i(t)$ in Eq.~(\ref{eq:nodalflowbal0}) stands for generally time-dependent amount of gas injected (positive) or consumed (negative) at the node $i$.
The second type of nodal boundary conditions states continuity of pressure at the joints
\begin{eqnarray}
\forall i,j,k\in{\cal V},\quad \mbox{s.t.}\quad \{i,j\},\{i,k\}\in {\cal E}:\quad p_{ij}(t,0)=p_{ik}(t,0).\label{eq:nodalpressure}
\end{eqnarray}

Gas flow through a sufficiently large network is normally controlled by compressors which are devices boosting pressure while preserving mass flow.
Typically compressors are placed at a pipe next to a joint, i.e. at the interface between a node and neigboring/adjoint pipes. We model compressor placed at the pipe $\{i,j\}$, next to the node, $i$, as a point of pressure discontinuity, thus enhancing inlet pressure, $p_i(t)=p_{ij}(t,0^{in})$ by a multiplicative and generally time dependent positive factor, $\gamma_{ij}(t)$, to
\begin{eqnarray}
p_{ij}(t,0^{(out)})=\gamma_{ij}(t)p_{ij}(t,0^{(in)}),
\label{eq:nodalcomp1}
\end{eqnarray}
where $0^{(in)}$ and $0^{(out)}$ denote inlet and outlet location of the compressor. We assume that $\gamma_{ij}(t)\geq 1$ if $\phi_{ij}>0$ and
$\gamma_{ij}(t)\leq 1$ otherwise.

To complete the statement of the Initial Boundary Value Problem (IBVP), governed by the system of PDEs \eqref{eq:mass3},\eqref{eq:momentum3} and the  boundary conditions
%\eqref{eq:nodalpressure},
\eqref{eq:nodalflowbal0}-\eqref{eq:nodalcomp1}, one also need to provide initial conditions at all points of the network at the initial time, $t=0$.

Let us also clarify (for completeness) that injections/consumptions at the nodes, $q_i(t)$, assumed known exogenously in the IBVP  formulation above, can also be replaced by fixed pressure condition, or a hybrid condition, at one or many nodes.

\section{Split-Step Method} \label{sec:method}

In this Section we present step-by-step description of the operator splitting or ``split-step'' method for pipeline simulation in the slow transient regime. The approach is similar to one that is widely used in numerical simulation of light wave propagation in fiber optics \cite{TA1984, Agrawal2001}.
We first review the concept of operator splitting, and then details its use to solve the system of hyperbolic PDEs.
Essence of the method consists in splitting the dynamics into two alternating steps.
The first step, representing dynamics of an auxiliary homogeneous linear hyperbolic system, is solved by propagation along characteristics of a space-time grid. Dynamics associated with the second step, represented by an auxiliary inhomogeneous nonlinear system, is spatially local. Both steps are designed to preserve the network boundary/compatibility conditions.

\subsection{Operator Splitting} \label{subsec:opsplit}

In our description of  the operator splitting technique we largely follow notations used in the fiber optics literature \cite{TA1984, Agrawal2001}.
Consider an evolutionary PDE for vector ${\bf y} = \left[\psi(t,x),\, \phi(t,x)\right]^T$ over a pipe stated in the following operator form:
\begin{align} \label{eq:split0}
{\bf y}_t & = (\hat L + \hat N) {\bf y}.
\end{align}
Here $\hat L$ and $\hat N$ denote the linear hyperbolic operator and the nonlinear D--W damping operator respectively.
Given a uniform time grid $0, \tau, 2\tau,\ldots,N\tau$ we denote by ${\bf y}^{(n)} = {\bf y}(n\tau)$.
Evolution of ${\bf y}$ over a time step $\tau$ is approximated through consecutive linear and nonlinear steps
\begin{align}
\tilde {\bf y}_{t} = \hat L {\bf y},\label{eq:split1a}\\
{\bf y}_{t} = \hat N\tilde {\bf y}. \label{eq:split1b}
\end{align}
%\begin{align}
%\tilde\psi^{(n+1)}_t = \hat L\tilde \psi^{(n+1)}, \mathrm{\;initial\; condition\;} \psi^{(n)},\label{eq:split1a}\\
%\psi^{(n+1)}_t = \hat N\psi^{(n+1)}, \mathrm{\;initial\; condition\;} \tilde\psi^{(n+1)}\label{eq:split1b}.
%\end{align}
%Stated explicitly, we apply \eqref{eq:split1a} to the previous state and then we apply \eqref{eq:split1b} to the result to obtain an approximation of
%evolution at time $t_{n+1}$.
The formal solution of \eqref{eq:split1a} and \eqref{eq:split1b} is written in terms of the operator exponent as follows:
\begin{align} \label{eq:split2}
\tilde{\bf y}^{(n+1)} & = \E^{\tau\hat L}{\bf y}^{(n)}, \qquad
{\bf y}^{(n+1)} = \E^{\tau\hat N}\tilde{\bf y}^{(n+1)}.
\end{align}
Solving Eqs.~\eqref{eq:split0} over a time step $\tau$ one derives
\begin{align}
{\bf y}^{(n+1)} &= \E^{\tau(\hat L + \hat N)}{\bf y}^{(n)}.\label{eq:split3}
\end{align}
Examining the error acquired in the $\tau\to0$  limit due to noncommutativity of the operators $\hat L$ and $\hat N$, one arrives at
\begin{equation} \label{eq:spliterror0}
\E^{\tau(\hat L + \hat N)}{\bf y}^{(n)} - \E^{\tau\hat L}\E^{\tau\hat N}{\bf y}^{(n)} = \left(\frac{\tau^2}{2}[\hat N, \hat L] + \mbox{h.o.t.}\right){\bf y}^{(n)},
\end{equation}
where $\mbox{h.o.t.}$ stands for higher order terms and $[\hat A, \hat B] = \hat A\hat B - \hat B\hat A$ is the commutator of $\hat A$ and $\hat B$.
The local error is $O(\tau^2)$ while the method is globally first-order in time. However, by using a ``symmetrized'' operator
\begin{align} \label{eq:symmetrized0}
{\bf y}^{(n+1)} = \E^{\frac{\tau}{2}\hat N} \E^{\tau\hat L} \E^{\frac{\tau}{2}\hat N}{\bf y}^{(n)}.
\end{align}
one improves the accuracy making the method globally second-order in time.
It is straightforward to show that the local error associated with this symmetrized approximation is given by
\begin{equation} \label{eq:symmetrizederror0}
\E^{\tau(\hat L + \hat N)}\psi^{(n)} - \E^{\frac{\tau}{2}\hat N}\E^{\tau\hat L}\E^{\frac{\tau}{2}\hat N}\psi^{(n)} \! = \! \left(\! \frac{\tau^3}{12}\left\{[\hat L,[\hat N, \hat L]] \!+\! \frac{1}{2}[\hat N, [\hat N,\hat L]]\right\} + \ldots\!\right)\!\psi^{(n)}.
\end{equation}
In summary, error of the symmetrized ``split-step'' method~\eqref{eq:symmetrized0} is locally $O(\tau^3)$, and it is $O(\tau^2)$ globally.

Notice that consideration above apply not only to the $\hat{L}$-linear and $\hat{N}$-nonlinear split, but also to any other split giving the same target, $\hat{N}+\hat{L}$ in result. However, in what follows we will stick to the outlined linear/nonlinear split.

%{\color{red} ... this is a placeholder for discussion of how we choose the splitting in the \eqref{eq:mass2}-\eqref{eq:momentum2} ... also is there a generalization which allows overlap between $N$ and $L$ ? ... } {\red KAO: Not that I'm aware of. Otherwise it wouldn't be operator splitting, right?}

\subsubsection{First Step: Evolution of the Linear Hyperbolic System} \label{subsec:linear}

The homogeneous part of \eqref{eq:mass3}-\eqref{eq:momentum3} is a first order system:
\begin{align}
& \dfrac{\d p}{\d t} + \dfrac{\d \phi}{\d x} = 0, \label{eq:hyper0a} \\
& \dfrac{\d \phi}{\d t} + \dfrac{\d p}{\d x} = 0, \label{eq:hyper0b}
\end{align}
and as such can be reduced to the second order wave equation and solved explicitly via the method of characteristics. Let us introduce the characteristic variables,
\begin{align}\label{eq:chartransform}
 W^{+} = \dfrac{p+\phi}{\sqrt{2}}, \quad  W^{-} = \dfrac{p-\phi}{\sqrt{2}}
\end{align}
which satisfy the transport equation
\begin{equation}
\label{eq:hyper2}
\begin{cases}
& W^{-}_t + W^{-}_x = 0,\\
& W^{+}_t - W^{+}_x = 0.
\end{cases}
\end{equation}
Observe that $W^{-}(x,t) = W^{-}(x-t)$ and $W^{+}(x,t) = W^{+}(x+t)$ is the exact solution of the homogeneous linear
system \eqref{eq:hyper0a}-\eqref{eq:hyper0b}. The values of pressure $p$ and flow $\phi$ are then given by:
\begin{align}
& p(x, t + \tau) = \dfrac{W^{+}(x+\tau, t) + W^{-}(x-\tau, t)}{\sqrt{2}} \label{eq:linupdate1}\\
& \phi(x, t + \tau) = \dfrac{W^{+}(x+\tau, t) - W^{-}(x-\tau, t)}{\sqrt{2}} \label{eq:linupdate2}
\end{align}
%{\red KAO: Sergey, I'm sorry, but $x+\tau$ looks like shit!}

Evolving~\eqref{eq:linupdate1} and~\eqref{eq:linupdate2} over one time step $\tau$ is the application of
the operator $\E^{\tau \hat L}$ to ${\bf y}$.

Restated in  characteristic variables
Eqs.~\eqref{eq:hyper0a}-\eqref{eq:hyper0b} turn into ${\bf y}_t = \hat L {\bf y}$ which is easy to solve over
rescaled variable regular space-time grid. In physical variables the resulting expressions are
\begin{equation}
\label{eq:Delta_t_Delta_x_relation}
c_s \Delta t = \Delta x.
\end{equation}
This formula requires introduction of inhomogeneous spacial grid along the pipe in order to keep the same $\Delta t$
at the varying $c_s$.

We conclude discussion of the first, linear, step, noticing that principally hyperbolic/characteristic approach described above allows generalization to the case of curved/nonlinear characteristics for an extra price of an additional interpolation step. (Accounting for curved characteristics will be needed, in particular, if a nonlinear advection term, dropped from the Euler equation in the basic model analyzed in this paper, is added to consideration.)

%{\color{red} ... placeholder for reference to other characteristics based methods used in the past for the gas-network modeling ... e.g. review of \cite{TT1987} ... and also linking it to the aforementioned "curved" characteristics ... is our choice of terms is based on the desire to keep it analytic?  are there other choices which are analytic ... but also include (more physical) balance of gradient of pressure and dissipation ... some part of this discussion may go into conclusions ...} {\red KAO: Misha, as far as I understand, our choice of terms is based on very simple philosophy: let's neglect as much as we can, keeping in mind that we need to get equations which are convenient for numerical solution. It happened so, that our choice has explanation from the point of view of physics (see comparison of terms).}

\subsubsection{Second Step: Evolution of Nonlinear Damping} \label{subsec:nonlinear}

We solve the nonlinear part of \eqref{eq:mass3}-\eqref{eq:momentum3} given by
\begin{equation}
\label{eq:odesys0}
\begin{split}
& \dfrac{\d p}{\d t} = 0, \\
& \dfrac{\d \phi}{\d t} = - \dfrac{f}{2D}\dfrac{\phi|\phi|}{p}.
\end{split}
\end{equation}
This equation can be integrated explicitly, which results in
\begin{align}  \label{eq:nlupdate}
 \phi(t+\tau) = \dfrac{\phi(t)}{1 + \alpha \tau \dfrac{|\phi(t)|}{p(t)}}
\end{align}
where $\alpha = f/D$. This exact solution of \eqref{eq:odesys0} is the application of
$\E^{\tau\hat N}$ to ${\bf y}$ in \eqref{eq:symmetrized0}.  Eq.~\eqref{eq:nlupdate}
is applied to $\phi$ at each point of the space grid independently of the others.

Aiming to derive conditions on the time-step $\tau$ let us analyze the error acquired in the result of approximating the operator $\hat N$ by $\E^{\tau\hat N}{\bf y}$. Definition of the operator exponent results in
\begin{align} \label{eq:opexponent}
\E^{\tau\hat N}{\bf y} = \left(\hat 1 + \tau \hat N + \frac{\tau^2}{2!}\hat N^2 + ...\right){\bf y}.
\end{align}
On the other hand, expansion of the right hand side of \eqref{eq:nlupdate} results in
\begin{align} \label{eq:opseries}
\phi(t'+\tau) = \phi(t') \left(1 - \alpha\tau\frac{|\phi(t')|}{p(t')} + \alpha^2\tau^2\frac{|\phi(t')|^2}{p(t')^2} + ...\right).
\end{align}
Comparing terms in Eqs.~(\ref{eq:opexponent},\ref{eq:opseries}) one finds out that the expressions become identical when the nonlinear operator is represented by $\hat N {\bf y}= [0,\,(-\alpha |\phi|/p)\phi]^T$.
Observe now that the series approximation \eqref{eq:opseries} of \eqref{eq:nlupdate} converges only if $\alpha\tau|\phi|/p < 1$, so that the
condition on $\tau$ for the approximation to be valid is
\begin{equation}\label{eq:nonlin_cond}
\tau < \frac{1}{\alpha} \min\left(\frac{p}{|\phi|}\right).
\end{equation}
We note that in practice, when a sufficiently small time step is chosen to yield reasonable accuracy of simulations, the condition (\ref{eq:nonlin_cond}) is satisfied with a large margin.

\subsection{Propagation of Characteristics Through the Network} \label{subsec:nodal}

\begin{figure}[htp!]
\centering
  \begin{subfigure}{\textwidth}
   \begin{center}
 \includegraphics[width = 0.45\textwidth]{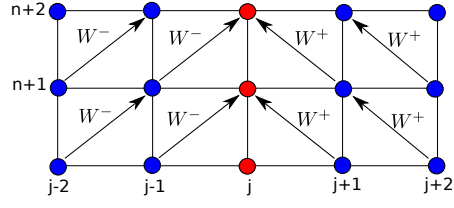}{\centering}
   \end{center}
  \caption{The values of pressure $p^{(n+1)}$ and mass flow variables $\phi_{in}^{(n+1)}$ and $\phi_{out}^{(n+1)}$ at the node depend on the values
  from two characteristics $W^{-}_{in}$ and $W^{+}_{out}$ from the incoming and from outgoing pipes, respectively.}
  \end{subfigure}
  \begin{subfigure}{\textwidth}
   \begin{center}
 \includegraphics[width = 0.45\textwidth]{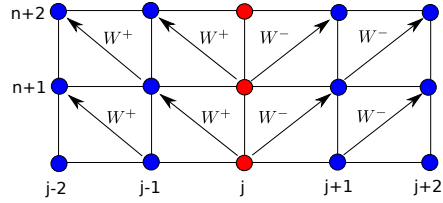}{\centering}
   \end{center}
  \caption{The values of $W^{-}_{in}$ and $W^{+}_{out}$ are supplied from the node to advance the values of pressure $p(x)$ and
mass flow variables $\phi_{in}(x)$ and $\phi_{out}(x)$ to the interior of incoming and outgoing pipes. }
  \end{subfigure}
 \caption{The horizontal direction is the spatial coordinate along the pipe and vertical direction is time. Red circles mark the position of a node that connects two pipes.}
 \label{fig:char2}
\end{figure}

The approach described in Sections \ref{subsec:linear} and \ref{subsec:nonlinear} provides sufficient base for devising numerical method to solve the system of Eqs. \eqref{eq:mass3}-\eqref{eq:momentum3} for a single pipe given the appropriate boundary and initial conditions.  The remainder of this Subsection is devoted to extension of the split-step method to the IBVP problem over the network model defined in Section \ref{sec:model}.

Notice that the second non-linear step of the split-step methodology, detailed in Section \ref{subsec:nonlinear}, is spatially local. This means that extension of this step to the case of a network is completely straightforward.

Extending the first linear step from the single-pipe case to the network case requires some extra work. Specifically, one needs, in addition to what was done in Section \ref{subsec:linear} for the graph-linear (pipe) elements of the network, to account for the mass flow balance at any node/junction governed by Eq.~(\ref{eq:nodalflowbal0}).

Let us first ignore compressors and write down the time-step incremental mass-flow balance at a node, $i$, relating consumption/production at the node, $q_i(t)$, to mass flows, $\phi_{ij}(t,0)$, from the node $i$ along the neighboring pipes, $\{i,j\}\in{\cal E}$:
\begin{eqnarray} \label{eq:eqmb}
q_i(t+\tau) = \sum_{j:\{i,j\}\in{\cal E}} \phi_{ij}(t,0) S_{ij}.
\end{eqnarray}
Then,  we replicate the single pipe characteristic Eqs.~\eqref{eq:hyper0a},\eqref{eq:hyper0b},  rewriting them for each line in the following form
\begin{eqnarray}\label{eq:w}
\qquad \forall j: \{i,j\}\in{\cal E}:\quad \phi_{ij}(t+\tau,0)=\left\{
\begin{array}{cc}
p_i(t+\tau)-\sqrt{2} W_{ij}^{-},&  \phi_{ij}(t+\tau,0)\leq 0\\
-p_i(t+\tau)+\sqrt{2} W_{ij}^{+},&  \phi_{ij}(t+\tau,0)>0
\end{array}
\right.
\end{eqnarray}
It is also useful for implementation (see the following Section \ref{subsec:algo}) to substitute $\phi_{ij}$ from Eq.~(\ref{eq:w}) into Eq.~(\ref{eq:eqmb}) thus rewriting the latter solely in terms of the $q$ and $W$ variables
\begin{eqnarray}\label{eq:pi}
\quad p_i(t)=\frac{\sqrt{2} \sum_{j:\{i,j\}\in{\cal E}}S_{ij}\left(\frac{W_{ij}^{-}+W_{ij}^+}{2}+
\mbox{sign}\left(\phi_{ij}(t)\right)\frac{W_{ij}^{-}-W_{ij}^+}{2}\right)-q_i(t+\tau)}{\sum_{j:\{i,j\}\in{\cal E}} S_{ij}}.
\end{eqnarray}
Overall, supplementing the system of Eqs.~\eqref{eq:linupdate1}, \eqref{eq:linupdate2}, stated for each pipe/line, by Eqs.~\eqref{eq:eqmb}, \eqref{eq:w}, stated for each node, complete description amenable for solution via the methods of characteristics.

Generalization of the linear step to the case involving compressors is also straightforward. Version of Eqs.~\eqref{eq:w}, \eqref{eq:pi} accounting for compression becomes ($\forall j: \{i,j\}\in{\cal E}$)
\begin{eqnarray}
\label{eq:w_comp}
\phi_{ij}(t+\tau,0)\!=\!\left\{\!
\begin{array}{cc}
\!\!\gamma_{ij}(t+\tau)p_i(t+\tau)\!-\!\sqrt{2} W_{ij}^{-},&  \!\!\!\!\phi_{ij}(t+\tau,0)\leq 0\\
\!\!-\gamma_{ij}(t+\tau) p_i(t+\tau)\!+\!\sqrt{2} W_{ij}^{+},&  \!\!\!\!\phi_{ij}(t+\tau,0)>0
\end{array}
\right.
\\
\label{eq:pi_comp}
p_i(t)=\frac{\sqrt{2} \sum_{j:\{i,j\}\in{\cal E}}S_{ij}\left(\frac{W_{ij}^{-}+W_{ij}^+}{2}+
\mbox{sign}\left(\phi_{ij}(t)\right)\frac{W_{ij}^{-}-W_{ij}^+}{2}\right)-q_i(t+\tau)}{\sum_{j:\{i,j\}\in{\cal E}} \gamma_{ij}(t+\tau)S_{ij}}.
\end{eqnarray}

The boundary conditions are set consistently with the method of characteristics so that no approximation is involved.
The resulting method is stable for all values of time step $\tau$. This allows the value of $\tau$ to be
determined solely based on the desired accuracy.

\subsubsection{Split-Step Algorithm} \label{subsec:algo}

Bringing together all the elements explained above we arrive at the following Algorithm.
\begin{enumerate}
\item[1a.] { Set the initial condition over the network: $\forall \{i,j\}\in{\cal E},\quad \forall x\in[0;L_{ij}]:\ \phi_{ij}(t=0,x) = f_{ij}(x)$ and $p_{ij}(t=0,x) = g_{ij}(x)$.}
\item[1b.] { Set initial pressures at all nodes of the network: $\forall i\in{\cal V}:\ p_i(t = 0) = R_i$. }
\item[2.]  { Apply one half nonlinear damping time-step at all the pipes:
\begin{equation*}
 \forall \{i,j\}\in{\cal E},\quad \forall x\in[0,L_{ij}]:\quad \tilde \phi_{ij}(t,x) = \dfrac{\phi_{ij}(t,x)}{1 + \alpha_{ij} \frac{\tau}{2} \dfrac{|\phi_{ij}(t,x)|}{p_{ij}(t,x)}}.
\end{equation*}
}
\item[3.] { Assign new values along the characteristics for all the pipes:
\begin{align*}
 \forall \{i,j\}\in{\cal E}, \forall x\in[0,L_{ij}]:\; & W_{ij}^{+}(t + \tau,x) = \dfrac{p_{ij}(t,x)+\tilde\phi_{ij}(t,x)}{\sqrt{2}} \\
 & W_{ij}^{-}(t + \tau,x) = \dfrac{p_{ij}(t,x)-\tilde\phi_{ij}(t,x)}{\sqrt{2}}
\end{align*}
}
\item[4.] { Apply full linear hyperbolic step at all interior points of all the pipes:
\begin{align*}
 \forall \{i,j\}\in{\cal E},\ \forall x\in[0;L_{ij}]:\quad & {\tilde p}_{ij}(t + \tau,x)    = \dfrac{W_{ij}^{+}(t,x+\tau) + W_{ij}^{-}(t,x-\tau)}{\sqrt{2}}\\
 & {\tilde\phi}_{ij}(t + \tau,x) = \dfrac{W_j^{+}(t,x+\tau) - W_j^{-}(t,x-\tau)}{\sqrt{2}}
\end{align*}
}
\item[5.] {
At all the nodes $i$ update pressure $p_i(t+\tau)$ and boundary flows according to
Eqs.~\eqref{eq:w_comp}, \eqref{eq:pi_comp}.
}
\item[6.]  {
Apply the second half of the nonlinear damping time-step at all the pipes:
\begin{equation*}
\forall \{i,j\}\in{\cal E},\quad \forall x\in[0,L_{ij}]:\quad \phi_{ij}(t+\tau,x) = \dfrac{{\tilde\phi}_{ij}(t+\tau,x)}{1 + \alpha_{ij} \frac{\tau}{2} \dfrac{|{\tilde\phi}_{ij}(t+\tau,x)|}{{\tilde p}_{ij}(t+\tau,x)}}.
\end{equation*}
}
\end{enumerate}
Repeat steps $2$ through $6$ until
%accuracy thresholds are exceeded or
the desired time is reached.

Note that both steps in \eqref{eq:symmetrized0} conserve mass of gas explicitly.  Therefore mass conservation is an intrinsic property of the
method,  and thus the accuracy of mass conservation does not depend on the size of the time step.

\section{Numerical Experiments and Comparisons} \label{sec:examples}

We present three case studies in which we demonstrate performance of our method and also compare it with the two
implicit methods.  We consider a single pipe, a single pipe with a compressor in the middle, and a small network of four pipes with a loop and
a compressor.  The simplest case of a single pipe will be used for comparison with the Kiuchi method \cite{Kiuchi1994}, and also to demonstrate
that the new method is capable of capturing acoustic waves propagating through the pipe as a transient resulting from a perturbation
containing sharp changes including all the harmonics.

\begin{figure}[htb]
 \begin{center}
 \includegraphics[width = 0.5\textwidth]{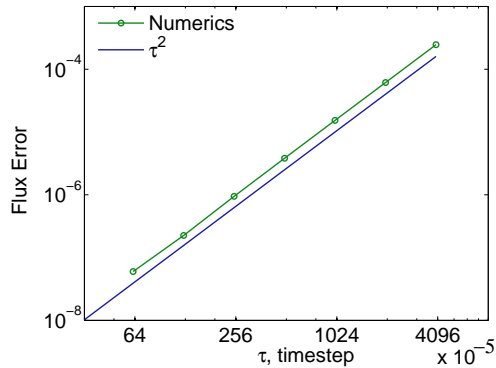}
 \end{center}
 \caption{\label{fig:ss2converge} Convergence of the operator splitting (split-step) method~\eqref{eq:symmetrized0} applied to~\eqref{eq:mass3}-\eqref{eq:momentum3} shows second-order
 scaling of global error with decrease of time step $\tau$ in complete correspondence with~\eqref{eq:symmetrizederror0}.
 %{\color{red} \sout{Need more information here.  What is the simulated example (i.e, pipe parameters and boundary conditions), and what flow error is examined?  Maybe just omit this?}}
Along vertical axis we show relative $L_2$ norm of the error in estimation of the flow against the ``numerically exact'' solution.} %The simplest one pipe case was considered.}

\end{figure}

First, we verify that, as stated in \eqref{eq:symmetrizederror0}, the global error converges to zero as $\tau^2$.
The Fig.~\ref{fig:ss2converge} confirms the second order convergence of the method.

\subsection{Single pipe} \label{sec:example1}

\begin{figure}
 \begin{center}
 \includegraphics[width = 0.49\textwidth]{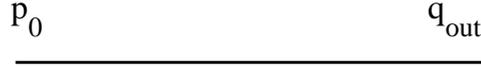}%one_pipe.eps
 \end{center}
 \caption{\label{fig:single_pipe_scheme}  Sketch of the test case of a single pipe.  The boundary conditions are inlet pressure $p_0$ and outlet flow $q_1$.}
% \label{ss2converge}
\end{figure}

The case of a single pipe, shown in Figure \ref{fig:single_pipe_scheme}, will be used to compare with the Kiuchi method \cite{Kiuchi1994}. The comparison aims to juxtapose stability of the methods, to evaluate the significance of terms in \eqref{eq:mass3}-\eqref{eq:momentum3}, and also to verify mass conservation. In this numerical experiment the inlet pressure and the outlet flow are fixed (constitute input), while the inlet flow and the outlet pressure are derived (constitute output).  (Notice, that the test is
mathematical rather than physical in nature because the nonlinear friction term in the phenomenological Weymouth Eqs. \eqref{eq:mass2}-\eqref{eq:momentum2} may require renormalization/adjustment in the case of fast changes in density/pressure. See conclusions for further discussions of these and other details of the physical modeling.)

We simulate a single pipe of length $L=20$ km and diameter $D=0.9144$ m ($36$ inches), with no compressors, speed of sound $c_s = 377.97$ m/s, and friction
factor $f=0.01$.  The simulation is initialized with a still gas (no flow) and the pressure of $p_0=6.5$ MPa kept uniform along the pipe.  Two different scenarios are considered. Scenario \#1 is highly dissipative with the perturbed waves damped fast. On the contrary scenario \#2 shows excitation and propagation of the weakly dumped waves.
%on the contrary waves e One case satisfies the necessary assumptions to use the approximation for rapid dissipation of waves.  The other case violates the approximations under which the dissipative term was derived, and indeed long lasting waves do occur.  We shall use the latter example in order to demonstrate the validity of keeping the time derivative term in the second equation and not including a nonlinear advection term there.

We choose to contrast performance of the split-step method with the Kiuchi's method \cite{Kiuchi1994}, because the latter is proven to be stable in the case of a single pipe.  The Kiuchi method is a finite difference implicit method, which remains stable even when the time-steps are large.  Fixed point iterations are used to resolve the implicit scheme.  Iterations are performed using the trust-region dogleg algorithm \cite{TrustRegion}.
Let us first of all mention that the Kiuchi method is advantageous over the split step method due to absence in the former of the rigid relation between the temporal and spatial steps, otherwise observed in the latter.

\subsubsection{Overdamped Waves} \label{sec:rapiddiss}

\begin{figure}
 \begin{center}
 \includegraphics[width = 0.49\textwidth]{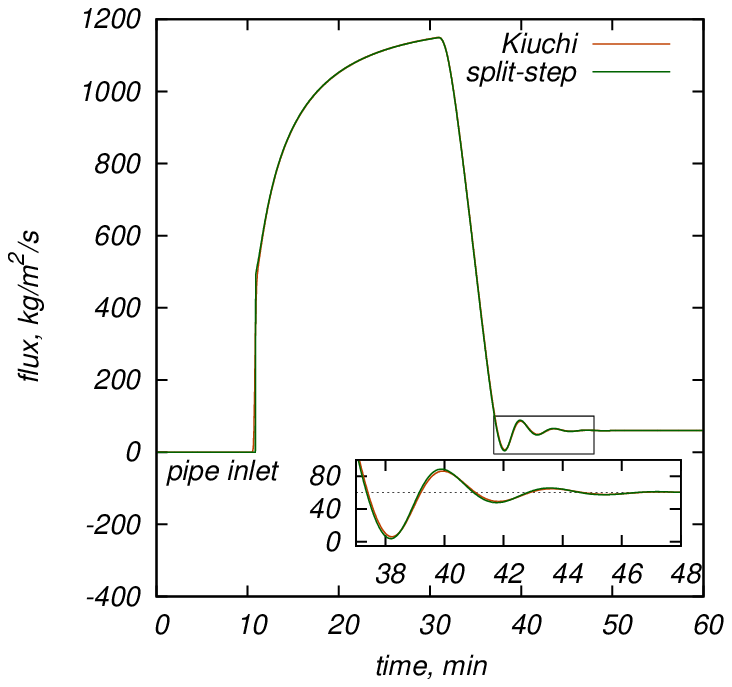}
\includegraphics[width = 0.49\textwidth]{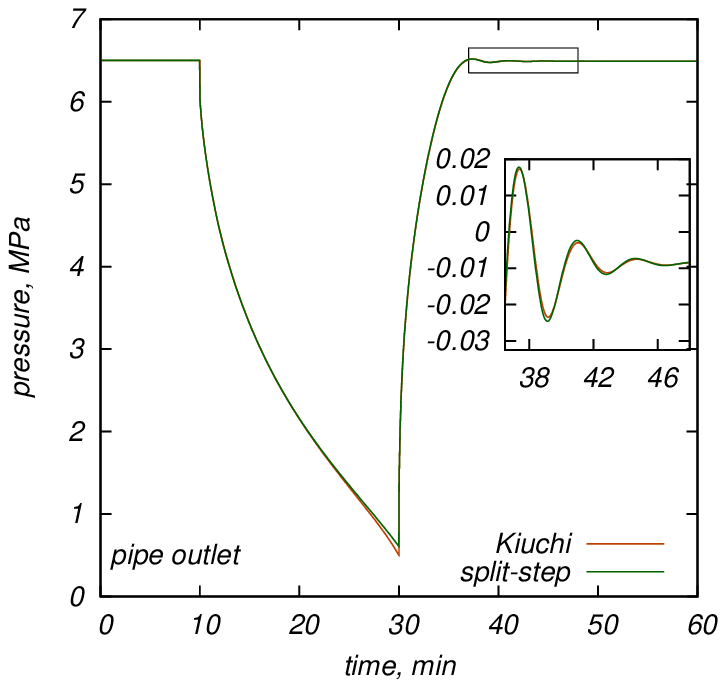}
 \end{center}
 \caption{\label{fig:Ex1b} Single pipe overdamped test. (Left) Flow at the left end of the pipe. (Right) Pressure at the right end
 of the pipe. Spatial step of $h=100$ meter and a time step of $\tau=1$ second is used in Kiuchi method implementation.
 Spatial step of $h=1.953$ meter and time step of $\tau=5.16743 \times 10^{-3}$ seconds is used in the split-step method implementation. At $t=30$ min the flow at the outlet (right) is reduced abruptly 20 fold.}
\end{figure}

Consider the following case:  at the time $t = 10$ min, the outlet is opened abruptly and gas is drawn at a rate $q_{\mathrm{out}}=\phi_0 S = 788.03$ kg/s.  Then at time $t = 30$ min, the consumption at the outlet is decreased  20 times and stays at this level for the remainder of the simulation experiment. The pressure is kept fixed to $p_0$ for the entire duration of the trial. The result of simulations using the operator-splitting method and Kiuchi's method are shown in Fig.~\ref{fig:Ex1b}.

One can observe oscillations in both pressure and flow in the pipe after the consumption was decreased. The reason for these oscillations is the formation
of a decaying standing acoustic wave along the pipe.  One can observe that dissipation of these waves happens during approximately one period of
oscillations. This corresponds to the situation where the dissipative term on the right hand side of in Eq.~\eqref{eq:momentum3} is dominant.
%, and the assumptions under which the dissipative term was derived are satisfied.

\subsection{Weakly Damped Waves} \label{sec:slowdiss}

\begin{figure}
 \begin{center}
 \includegraphics[width = 0.49\textwidth]{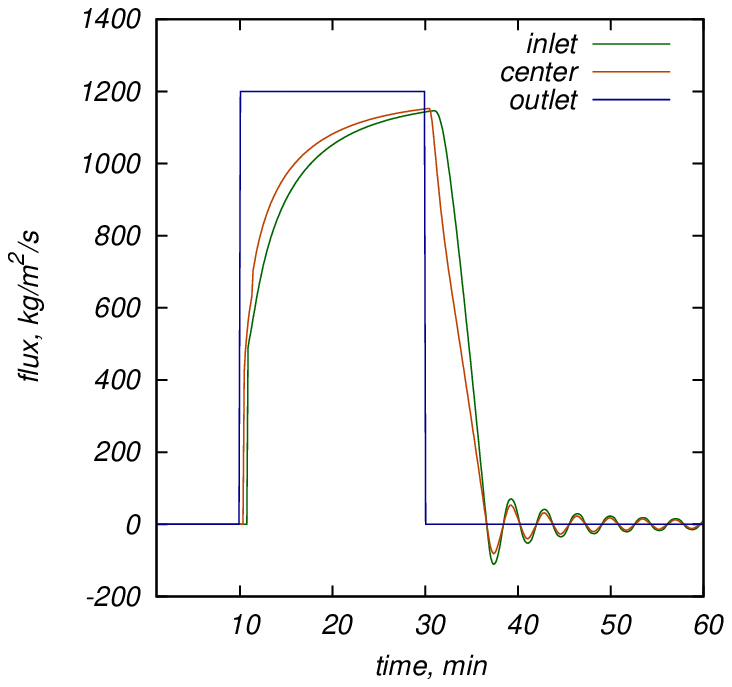}%{gas_flow_1.eps}
\includegraphics[width = 0.49\textwidth]{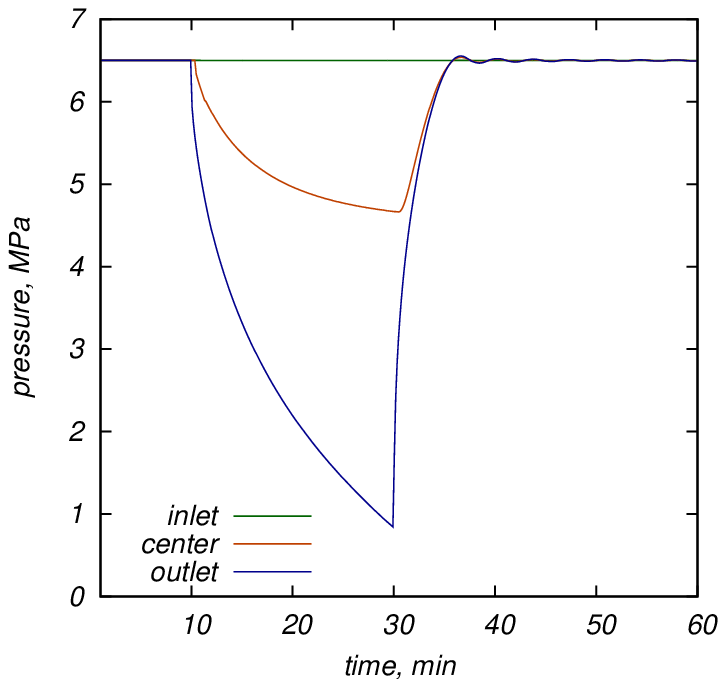}%{pressure_1.eps}
 \end{center}
 \caption{\label{fig:single_pipe_pressure_flow} Single pipe test illustrating formation of an almost (weakly dumped) standing wave. (Left) Flow at endpoints
 and midpoint of the pipe. (Right) Pressure at at endpoints and midpoint of the pipe. At $t=30$ min the flow at outlet (right end-point) is reduced to zero, otherwise parameters of the test case are equivalent to these explained in Section
 \ref{sec:rapiddiss} and illustrated in Fig.~\ref{fig:Ex1b}.}
% \label{ss2converge}
\end{figure}

Here we discuss split-step simulations in the case when the effect of nonlinear damping is weaker than in the simulations discussed in the preceding Subsection.
Specifically,  instead of reducing the outlet consumption by factor of 20 we block consumption completely at $t=30$ min.
Simulation results using the operator-splitting method are shown in Fig.~\ref{fig:single_pipe_pressure_flow}.  As the flow
approaches zero, the dissipative term decreases resulting in a sustainable (weakly damped) wave regime. We observe oscillations similar to the case of a standing wave in the pipe, i.e. ones observed when the dissipation is completely ignored. In this (standing wave) case period of oscillations can be estimated
from the respective linear approximation also taking into account proper boundary conditions (specifically $\phi_x(t,0)=0$ and $\phi(t,L)=0$ at $t\geq 30$ min). We observe (weakly decaying) standing wave with a quarter of a full period of a sinusoidal curve.
This means that the length of the full period of the wave would
be four times longer than the length of the pipe. As a result we can estimate time period of the oscillations as $T=4L/c_s \simeq 3.5$ min.
Fig.~\ref{fig:single_pipe_pressure_flow} shows that there are a slightly less than three oscillations per 10 min period, which is in a close
agreement with the estimation above.

Notice that numerical simulations discussed above are of a synthetic test type aimed to test the quality of our newly developed split-step method. However, this
underdamped regime may not be of a practical relevance, as violating physical conditions
used to estimate the D-W dissipative term. Indeed the D-W estimations are justified only in the turbulent regime when the flows are sufficiently strong. The dissipative term should be modified, possibly through a phenomenological interpolation between turbulent and laminar regimes,  which is beyond the focus/subject of this manuscript.

\subsubsection{Significance of Terms within the momentum balance equation} \label{sec:terms}

%We can use the single pipe case study to estimation the relative significance of the terms in \eqref{eq:mass3}-\eqref{eq:momentum3}.

As explained in Section \ref{sec:physics} we transition from the basic momentum equations from Eq.~(\ref{eq:momentum1}) to Eq.~(\ref{eq:momentum3}), dropping the self-advection term as small, however we keep in our simulations the time-derivative term in Eq.~(\ref{eq:momentum3}),  even though some of the approximation methods, noticeably  \cite{HMS2010,osiadacz84}, argue that ignoring the time-derive term would also be legitimate.

\begin{figure}[htb]
 \begin{center}
 \includegraphics[width = 0.49\textwidth]{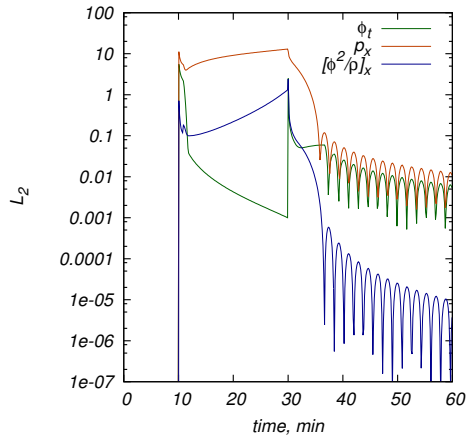}
 \end{center}
 \caption{\label{fig:terms_comparison} $L_2$ norms of the terms from Eq.~(\ref{eq:momentum3}), as observed in simulations of
  Eqs.~(\ref{eq:mass3},\ref{eq:momentum3}) by the split-step methods in the setting of the single-pipe experiment discussed in Section \ref{sec:slowdiss}.   }
\end{figure}

In order to verify significance of keeping the dynamic term (first term on the left hand side of Eq.~\eqref{eq:momentum3}) and the dropped self-advection term (second term on the left hand side of Eq.~\eqref{eq:momentum1})  we compare the $L_2$-norms of these terms with the pressure gradient term (last term on the left hand side of Eq.~\eqref{eq:momentum3}). The comparison is shown in Fig.~\ref{fig:terms_comparison}. Recall that we expect that the pressure gradient term mainly balances the nonlinear term on the right hand side of Eq.~\eqref{eq:momentum3} in the stationary regime and also in a slowly evolving regime.

Observe that during the slow process of gas flowing from the open valve between $t= 10$ min and $t= 30$ min, the pressure gradient term, shown red in Fig.~(\ref{fig:terms_comparison}), dominates both the flow derivative term (green) and the self-advection term (blue). The situation changes after we shut the valve off and generate an (almost) standing wave. While the self-advection term is still significantly smaller than the main term, the dynamic term and the main (pressure gradient) term become comparable. In this oscillatory (almost standing wave) regime the dissipative term (not shown in Fig.~(\ref{fig:terms_comparison})) is much smaller.

Hence, one can see that if acoustic waves are generated it makes sense to keep the time derivative of the flow while still neglecting the self advection term, as done in transition from Eq.~(\ref{eq:momentum1}) to Eq.~(\ref{eq:momentum3}). In other words, our model provides a reasonable bridge between simulation of full Euler equations and
the commonly accepted quasi-static equations \cite{HMS2010,osiadacz84}.

What we also learn from this comparison is that even though the pressure gradient term remains dominant even during abrupt changes (as occurred at $t=10$ min and following transient in the simulations discussed) the self-advection term, ignored in our simulations, becomes comparable to the dynamic term (kept in our simulations). This suggests that extending the split-step method to account for the self-advection term may be of interest for more accurate modeling of the cases including abrupt changes and following transients. (See concluding Section \ref{sec:conc} for further discussions of future work.)

\subsubsection{Test of mass conservation}  \label{sec:massconserv}

We aim to analyze the mass conservation quality within the operator splitting method, and also to contrast it with the mass conservation quality of the Kiuchi method. To simplify the mass conservation test one  prepares custom initial condition satisfying the following desired properties. First, of all one would like to have
smooth periodic solution in order to exploit spectral accuracy of trapezoid method for numerical integration, and therefore excluding respective numerical errors.) The trapezoidal rule for smooth periodic function is spectrally
accurate~\cite{Ryabenkii2000}, thus the total mass $M(t)$ can be computed to double-precision
with the following simple formula:
\begin{align}\label{eq:mass}
M(t) = \int\limits_0^L \rho(t,x) S\,\D x = \int\limits_0^L \frac{p(t,x)}{c_s^2} S\,\D x \approx \frac{Sh}{c_s^2}\left(\frac{1}{2}p(t,x_0) + \sum_{j=1}^{N-1} p(t,x_j) + \frac{1}{2}p(t,x_N)\right),
\end{align}
where $S$ is the pipe cross-section and $h$ is the elementary spacing along the pipe. One
considers mass values derived with this method to be ``numerically exact''.
%, i.e. we neglect an error associated with numerical integration.
Second, one chooses initial condition which is sufficiently far from a stationary solution of the simulated Eqs.~\eqref{eq:mass3},\eqref{eq:momentum3}. This is to guarantee that the initial condition show in parallel with the mass conservation some change in time, i.e. non-trivial dynamics, for the flux and pressure spatio-temporal profiles.
%Second requirement/property for the test initial condition is not to satisfy to be far enough from stationary solution. If we start from the stationary solution, Kiuchi method, as an iterative one, never depart from it, because iteration convergence condition (relative difference between two successive iterations) will always be satisfied, as a result solution will not be changed which will give us perfect conservation; while operator-splitting method, as an explicit one, will always be changing the solution.
%As stated in the introduction, we consider our method covering the gap between methods suitable for situations close to the balanced almost stationary flow and methods for extremely fast transients. As a result we need to have some dynamics which will allow us to estimate mass conservation errors introduced by numerical methods.

\begin{figure}
\begin{center}
 \includegraphics[width = 0.49\textwidth]{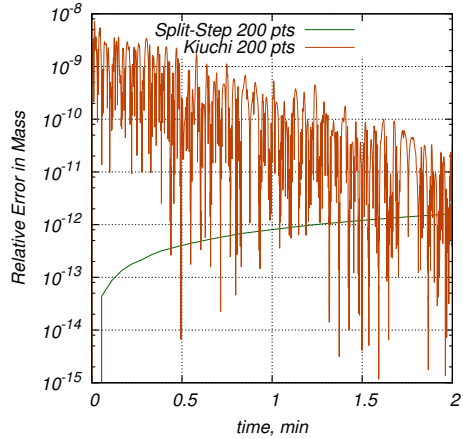}%
 \caption{\label{mass_comp}  Comparison of errors in the mass conservation seen in the test case (see description in the text) with the operator-splitting and Kiuchi methods run with the same time discretization and error-tolerance.
 %Even with an error tolerance of $10^{-14}$, the Kiuchi method provides lower accuracy in mass conservation at the beginning, when transient processes take place.
 }
 \end{center}
\end{figure}

A synthetic setup satisfying the two requirements is as follows. One uses a pipe of length $L=20$ km, of diameter $D=0.9144$ m, and of the friction coefficient $0.01$.  The speed of sound in the gas is taken as
$c_s = 377.9683$ m/s.  The initial conditions include stationary initial flow $\phi(0,x)=0$ along the pipe, with both intake
and output valves shut completely so that the boundary conditions are $\phi(t,0)=\phi(t,L)=0$.   The initial
pressure distribution is given by $p(0,x) = 6.5 + 0.2\cos(16\pi x/L)$ MPa. Notice that, admittedly synthetic (designed for test only) IBVP yields (decaying) periodic solution for
$p(x,t)$ and $\phi(x,t)$ which remain smooth in $x$ at all times, $t$. We track error accumulation of the split-step and Kiuchi methods and show the results in Fig.~\ref{mass_comp}.

Examination of Fig.~\ref{mass_comp} suggests that in the case of a significant transient dynamics (early in the test) error of the Kiuchi method is at least by an order of magnitude larger than of the split-step method. Later in time, when turbulent friction
brings solution closer to a stationary state mass conservation error of the Kiuchi method improves and become comparable to these of the split-step method. This improvement of the Kiuchi method performance when solution approaches a stationary solution is expected.

%Essentially, the closer we are to the stationary solution, the better Kiuchi method conserves total gas mass in the pipe. This behavior is what one should expect from an iterative method applying the logic we used choosing the initial conditions with dynamic transient.

Evaluating performance of the split-step method one observes that the mass conservation error per step is of the order of the round off error for double precision. However this extremely small error accumulates in time, which is seen in the slow linear growth of the operator-splitting curve in Fig.~(\ref{mass_comp}) and can be explained as follows. According to Eq.~(\ref{eq:mass}) time derivative of the total mass is
\begin{align}
\frac{\D M}{\D t} = \int\limits_0^L \frac{\pD \rho}{\pD t} S\,d x = - \int\limits_0^L \frac{\pD (\rho u)}{\pD x} S\,\D x \ge C \sqrt{N}\varepsilon,
\end{align}
where $N$ in the number of grid points along the pipe, $\varepsilon$ is the roundoff error of finite precision arithmetics, and $C$ is a $O(1)$ constant dependent on the maximum value of flow in the pipe.
Indeed, one observes  that since all the integrations in Eq.~(\ref{eq:mass}) are performed numerically, the last integral, which would be zero in exact mathematics, results in  an error accumulation because of the finite precision of computer arithmetics.

In summary,  the simulations confirm an important and desirable property of the operator-splitting method -- intrinsic mass conservation in particular in the regimes with significant pressure and flux transients.
%It should be noted, that in the case of slow transients, when a solution is almost stationary, Kiuchi, as well as other iterative methods may be tuned, on the expense of time-step decrease, to show comparable performance. The strong side of the Kiuchi method for slow dynamics is possibility to take long time step, while in split-step due to exact solution of the linear parts of equations requirement for the characteristics to pass exactly through grid points results in rigid relation~\eqref{eq:Delta_t_Delta_x_relation}.

\subsection{Pipe with compressor \& test of causality} \label{sec:example2}

Here we discuss a model including compressor with a time-varying compression ratio.  We also test how the split-method handles causality associated with speed-of-sound propagation of changes along the pipe.

\begin{figure}
 \begin{center}
  \includegraphics[width = 0.49\textwidth]{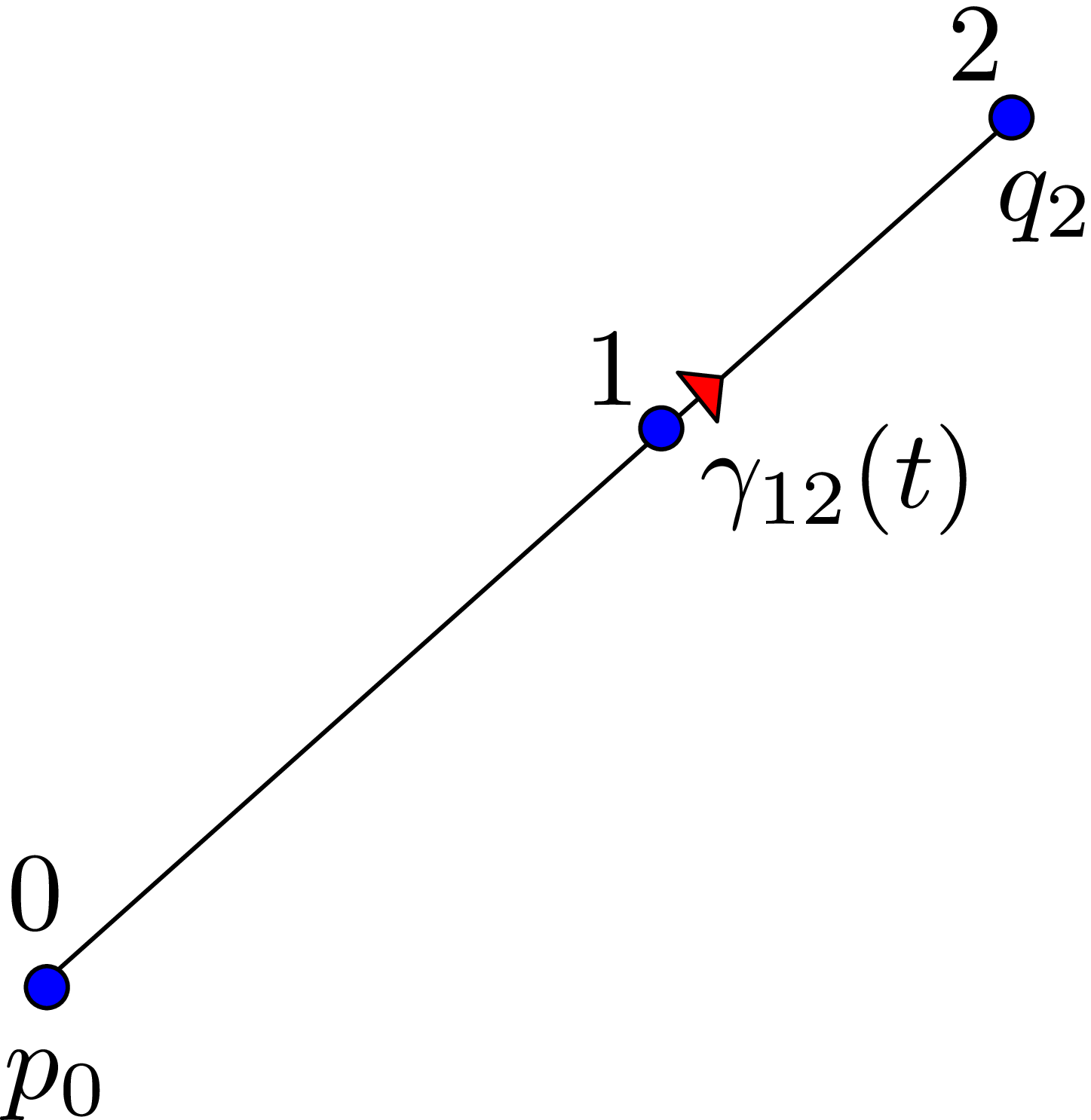} \includegraphics[width = 0.5\textwidth]{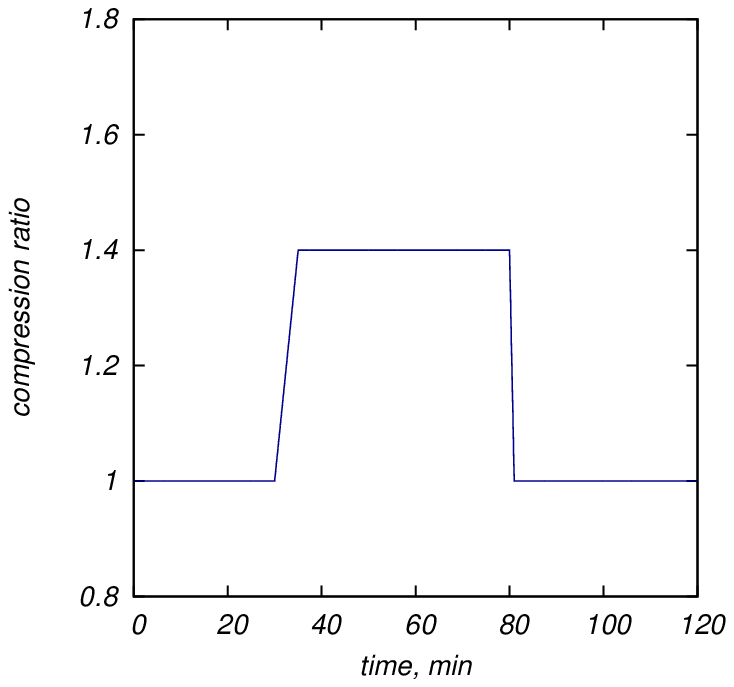} %pipe_with_compressor.eps
 \end{center}
 \caption{\label{fig:compression_ex2} Left: Layout of a single pipe with compressor. The inlet
 pressure $p_0$ and outlet flow $q_2$ are fixed. Compressor with the ratio $\gamma_{12}$ is located next to node 1. Right: compression ratio $\gamma_{12}(t)$ as a function of time chosen for the test case.}
\end{figure}

Consider a pipe with compressor as shown in Fig.~(\ref{fig:compression_ex2}).  With the network notation used in
Section \ref{sec:model}, the system parameters are
\begin{align}
&L_{01} = 50\,km\quad D_{01} = 0.9144\,m,\\
&L_{12} = 20\,km\quad D_{12} = 0.9144\,m,
\end{align}
the speed of sound is $c_s = 377.9683\,m/s$, and the friction factor is $f_{01}=f_{12} = 0.01$. Initial data for two
pipes is the steady state solution with constant steady
flow $\phi_{ij}$ and steady pressure $p_{ij}$:
\begin{align}
\phi_{ij}(0,x) = \phi_{ij}^0, \quad
p_{ij}(0,x) = \sqrt{p_i^2 - \dfrac{f_{ij} c_s^2}{D_{ij}} \phi_{ij}^0|\phi_{ij}^0| x},
\end{align}
where $x$ increases from 0 to $L_{ij}$ along the pipe in the direction from $i$ to $j$, and $p_i$ denotes initial
pressure at node $i$.  The following parameters are used to construct this initial state:
\begin{align}
&\phi_{01}^0 = 320, \quad \phi_{23}^0 = 320\, kg/m^2/s \\
&p_1 = 6.5\quad MPa  \\
&p_2 = \sqrt{p_1^2 - \dfrac{f_{01} c_s^2}{D_{01}} \phi_{01}^0|\phi_{01}^0| L_{01}}\quad MPa, \\
&p_3 = \sqrt{p_1^2 - \dfrac{f_{12} c_s^2}{D_{12}} \phi_{12}^0|\phi_{12}^0| L_{12}}\quad MPa
\end{align}

The compression ratio is a continuous, piecewise-linear function of time with a transition of the compression ratio from
off (at $\gamma_{12}(30)=1$) to on (at $\gamma_{12}(40) = 1.4$) over $\tau = 10$ minutes and back to off again within $\tau = 2.5$ minutes, as shown in Fig.~\ref{fig:compression_ex2}, and as described below:
\begin{equation}
\gamma_{12}(t) = \begin{cases}
 1\quad &\mbox{for}\, 0\leq t\leq 1800\, \mbox{sec.} \\
 -1.4+\frac{4}{3000}t \quad &\mbox{for}\, 1800 \leq t \leq 2100\, \mbox{sec.} \\
 1.4 &\mbox{for}\,  2100 \leq t \leq 4800\, \mbox{sec.} \\
 33.4-\frac{2}{300}t \quad &\mbox{for}\, 4800 \leq t \leq 4860\, \mbox{sec.} \\
  1 &\mbox{for}\,  4860 \leq t \leq 7200\, \mbox{sec.}
\end{cases}
\label{compression_ex2_formula}
\end{equation}
The gas withdrawal from node $i$ is given by $q_i(t)$ in kg/s as
\begin{align}
q_0(t) &= 0, \qquad q_1(t) = 0, \qquad q_2(t)= f_{01}.
\end{align}
For this example we compare results of simulations using the proposed method and a spatially lumped-temporarily orthogonal decomposition based  implicit scheme (we will call it just lump in the following), that has been recently developed for coarse-grained simulation and optimal control of gas pipeline networks \cite{ZCB2015}.  The lumped element solution is implemented using backward differentiation of variable
order (1 to 5) depending on the prescribed tolerance. %(we used solver \texttt{ode15i} in MATLAB)}.
\begin{figure}
 \begin{center}
 \includegraphics[width = 0.49\textwidth]{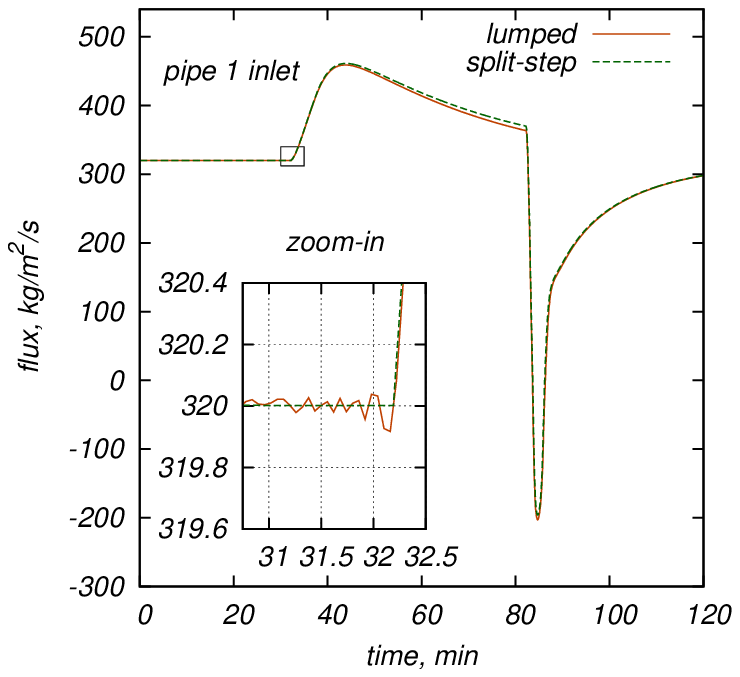}%{gas_flow_2.eps}
\includegraphics[width = 0.49\textwidth]{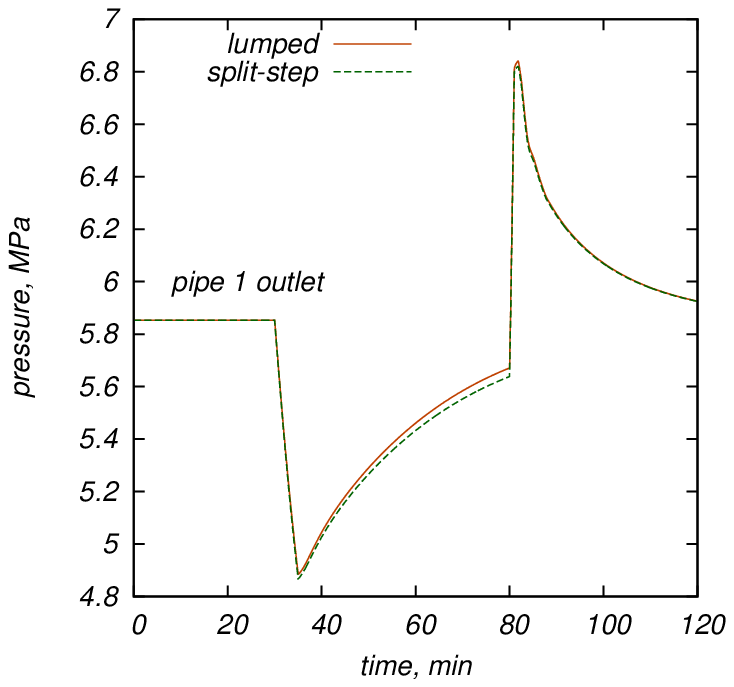}%{pressure_2.eps}
 \end{center}
 \caption{\label{fig:compressor_comp}Comparison of the operator-splitting method with the lumped element implicit
 scheme \cite{ZCB2015}.
 The operator-splitting solution is implemented using a space discretization of $h=1.953$ meters and time step of $\tau=5.16743 \times 10^{-3}$ seconds.}
\end{figure}

%The results of comparison of the proposed method with lumped element scheme~\cite{ZCB2015} is illustrated in Fig.~\ref{fig:compressor_comp}. We observe flow oscillations in time (shown in zoom-in) at the pipe inlet just after 30 minutes. However, the information about the new compressor state is transported $50$ kms by acoustic waves at the speed of sound and take $2.21$ minutes to reach the inlet: the effect is expected to reach inlet at time $T = 32.21$ mins. The oscillations exhibited by lumped element method before $T$ violate causality and persist on the timescale of minutes. The peak amplitude of oscillations is negligible preceding an almost discontinuous transition of the compression ratio from $c_1 = 1.0$ to $c_2 = 1.4$ and is followed by a smooth pressure/flow state after the transition. However, the smal relative difference in gas flow is amplified by at least an order of magnitude after the transition. For the case of instant compressor state switching, the difference between the two methods is expected to be more pronounced.

The comparison of the proposed method with the lumped element scheme~\cite{ZCB2015} is illustrated in Fig.~\ref{fig:compressor_comp}, and by inspection the results practically coincide.

As a nuance which may be important for studies sensitive to details of how exogenously excited changes transfer along the pipe, note that information about the new compressor state is actually transported $50$ km by acoustic waves at the speed of sound, which requires $2.21$ minutes to reach the inlet, so a change in inlet flow is expected to occur at time $T = 32.21$ mins.  For the split-step we observe steady flow (shown in zoom-in) at the pipe $1$ inlet just after 30 minutes.  In contrast, the lumped element solution (as not designed to conserve the causality exactly) shows oscillations before the moment of time $T$ thus violating causality, also persisting on the timescale of minutes.  The peak amplitude of oscillations is negligible, however, the small relative difference in gas flow is amplified by at least an order of magnitude after the transition.

%Numerical methods based on polynomial series expansion (e.g. Taylor series), such as Crank-Nicholson finite difference stencil or Kiuchi method exhibit oscillation or dissipative damping at the points of discontinuity in the function or its derivatives. This is typically not a problem for a hyperbolic system with dispersion or dissipative damping. On the other hand, the Darcy--Weisbach damping model does not introduce neither dispersion, nor dissipative damping. As a result discontinuities present in the initial conditions, or through rapid changes of the boundary conditions may remain in the solution for long times.

Two methodological comments following from these test numerical experiments are in order.

Numerical methods based on polynomial series expansion (e.g. Taylor series) such as the Crank-Nicolson finite difference stencil result in 
oscillation or dissipative damping at the points of discontinuity in the function or its derivatives. 
%Implicit methods such as the  Kiuchi method or the lumped element scheme will exhibit numerical errors in the implicit fixed-point iteration at each time step.  
While this is typically not a problem for a hyperbolic system with dispersion or dissipative damping, the Darcy--Wiesbach damping 
model does not introduce either dispersion or viscous dissipative damping, and as a result discontinuities present in the initial conditions, 
or caused by rapid changes in the boundary conditions, may persist in the solution for long times.

%Because ``split-step'' uses method of characteristics to solve the linear hyperbolic system, 
%it has the advantage of exactly replicating the causal properties of Weymouth equations. 
%Although treatment of discontinuous solutions is outside of their applicability scope, ``split-step'' 
%is capable of simulating discontinuous initial/boundary data, including accurate propagation of shock wave 
%through the gas network (in Weymouth framework, but not in the primordial Euler equations). This can be 
%important when simulating long time evolution (e.g. hours to weeks), when the timescale for the compressor 
%state transition (seconds to minutes) is comparable to the size of the numerical timestep and thus instant 
%compressor state switching is desirable. As demonstrated in the figure, instant compression state transition 
%poses no difficulty for the ``split-step'' method.

Because the ``split-step'' approach uses the method of characteristics to solve the linear hyperbolic system, 
it has the advantage of exactly replicating the causal properties of Eqs.~\eqref{eq:mass3}-\eqref{eq:momentum3}.  
Although these equations are not strictly applicable to the treatment of discontinuous conditions, the ``split-step'' 
method is capable of simulating the effects of discontinuous initial/boundary data, including accurate propagation of 
shock waves or other instantaneous changes through a gas pipeline network. 
 In particular, when long time evolution (e.g. hours to weeks) is of interest, it becomes reasonable to model 
instant switching of compressor states, since the timescale of compressor state transition (seconds to minutes) is 
of the same order, or smaller than the numerical timestep.  As demonstrated in the figure, instant state transition 
poses no difficulty for the ``split-step'' method.

\subsection{Simple Network With Joints and Cycles} \label{sec:example3}

\begin{figure}
 \begin{center}
 \includegraphics[width = 0.49\textwidth]{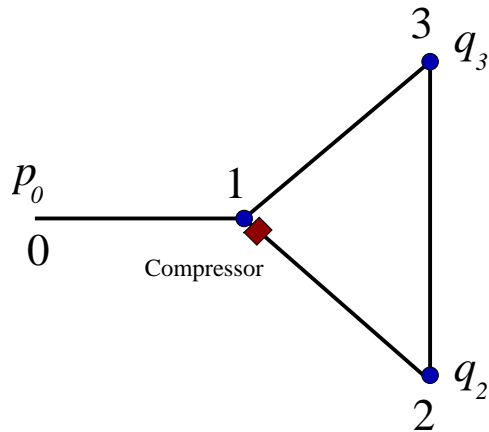} %
 \end{center}
 \caption{Schematic of a simple network test case. }
 \label{fig:simple_network}
\end{figure}

Finally, we present an example of a simulation on a simple pipeline network with a loop, which is chosen to emphasize that the split-step as an explicit method produces a consistent and accurate simulation for meshed networks.  The structure of the test network is shown in Fig.~\ref{fig:simple_network}.  The following parameters were used:
\begin{align}
&L_{01} = 50\,km, L_{12} = 80\,km, L_{13} = 80\,km, L_{23} = 80\,km, \\
& D_{01} = D_{12} = D_{13} =  0.9144 \,m, \quad  D_{23} = 0.6355\,m, \\
& f_{01} = f_{12} = f_{13} = f_{23} = 0.01,
\end{align}
with sound speed $c_s = 377.9683\,m/s$ for all the pipes. The system is initialized in the steady state solution with constant flow
$\phi_{ij}$ and steady pressure $p_{ij}$:
\begin{align}
\phi_{ij}(0,x) = \phi_{ij}^0, \quad
p_{ij}(0,x) = \sqrt{p_i^2 - \dfrac{f_{ij} c_s^2}{D_{ij}} \phi_{ij}^0|\phi_{ij}^0| x}.
\end{align}
Remind, that according to our notations, $x$ increases from 0 to $L_{ij}$ along the pipe in the direction from $i$ to $j$, and $p_i$ denotes initial pressure at node $i$.  The following setting is used to construct the initial state:
\begin{align}
&\phi_{01}^0 = 320, \quad \phi_{12}^0 = \phi_{13} = 160, \quad \phi_{23}^0 = 0\, \mathrm{kg/m}^2/\mathrm{s} \\
&p_0 = 6.5\quad \mathrm{MPa}  \\
&p_1 = \sqrt{p_0^2 - \dfrac{f_{01} c_s^2}{D_{01}} \phi_{01}^0|\phi_{01}^0| L_{01}}\quad \mathrm{MPa}, \\
&p_2 = \sqrt{p_1^2 - \dfrac{f_{12} c_s^2}{D_{12}} \phi_{12}^0|\phi_{12}^0| L_{12}}\quad \mathrm{MPa}, \\
&p_3 = \sqrt{p_1^2 - \dfrac{f_{23} c_s^2}{D_{13}} \phi_{13}^0|\phi_{13}^0| L_{13}}\quad \mathrm{MPa}.
\end{align}
For the boundary condition we choose to fix pressure at node $0$, $p_0 = 6.5$ MPa, and study the following injection/consumption temporal profiles at the other nodes (here and below the flows are measured in kg/s)
\begin{align}
q_1(t) & = 0, \quad q_2(t) =  96 + 64\cos(40\pi t/96), \quad q_3(t) =  160 + 96\sin(80\pi t/96)
\end{align}
The compression ratio of the compressor controlling pressure at the interface of node 1 and pipe $(1,2)$ is given by
\begin{align}
\gamma_{12}(t) & = 1 + \frac{1}{2}e^{-\kappa(t-t_c)^2},
\end{align}
where $t_c = 12$ hours and $\kappa = \frac{3}{40}$ 1/hour.

\begin{figure}
 \begin{center}
 \includegraphics[width = 0.49\textwidth]{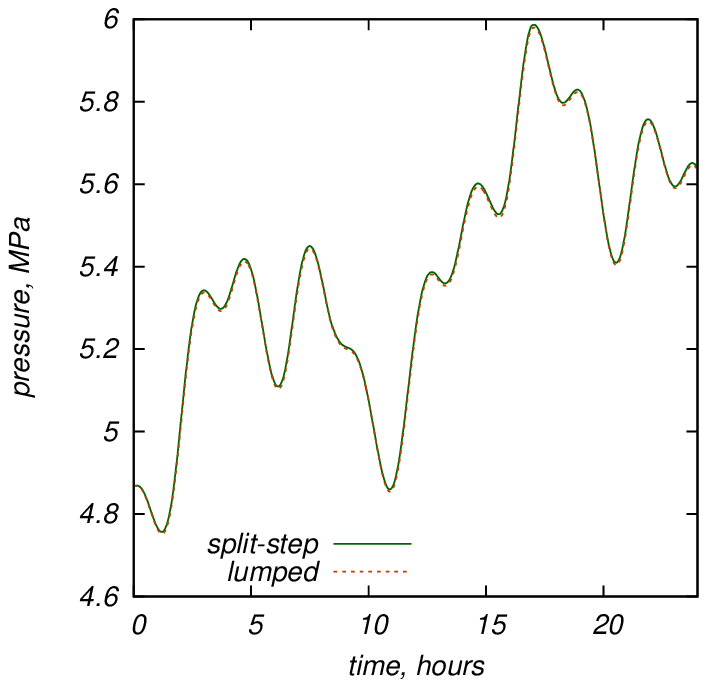}
 \includegraphics[width = 0.49\textwidth]{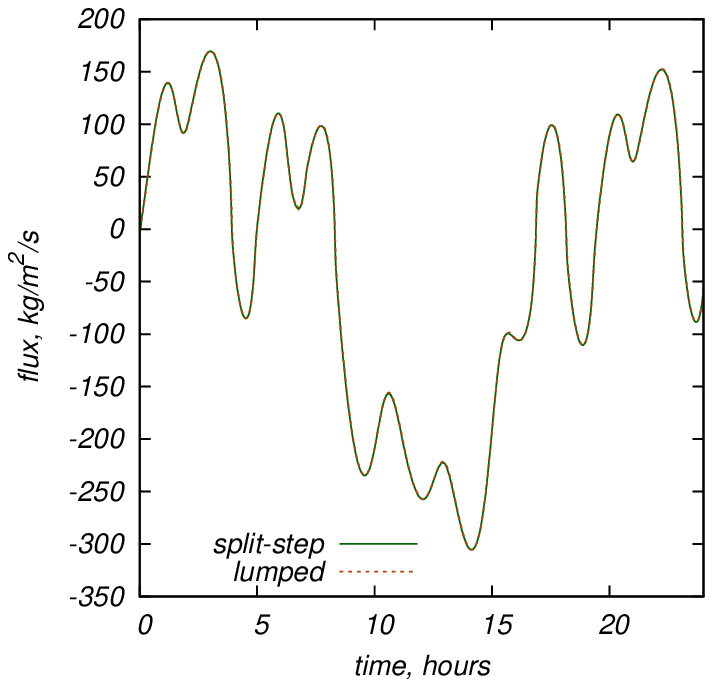}
 \end{center}
 \caption{\label{fig:simple_network_pressure_flow}
 Comparison of simulations for a small network example using the  explicit split-step method
 and the implicit lumped element method. Left: pressure at node $1$; Right: flow at node $2$ along the edge $(2,3)$. The operator-splitting method uses spatial discretization of $h=31.25$ meters and time step of $\tau=8.2679 \times 10^{-2}$ seconds.
 %We observe that in the case of initial and boundary conditions that create smooth transients without discontinuities, the solutions are indistinguishable for practical purposes.
 }
\end{figure}

Numerical solution of this IBVP for the period of $T=24$ hours is illustrated in Fig.~\ref{fig:simple_network_pressure_flow},
showing pressure at node 1 (the compressor inlet) and mass flow at the node $2$ along the $(2,3)$ pipe.
Notice good stability which might of been broken in the network with a loop as an expectation of an explicit method handicap.
We observe that solutions obtained using the explicit operator-splitting method and the implicit lumped  method are practically indistinguishable.

\section{Conclusion \& Path Forward} \label{sec:conc}

This paper was devoted to analysis of gas flows and pressure dynamics over natural gas networks modeled via the Weymouth system of differential equations~\cite{Osiadacz1989}, subject to time varying and generally unbalanced injection/consumption as well as time-varying compression.

Main result of the paper consists in adopting the operator-splitting method \cite{strang68,TA1984,Agrawal2001} to modeling pressure and mass flow dynamics over gas networks in a variety of regimes \cite{wylie78,thorley87,hudson06} ranging from slow dynamics, governed by a spatially local balance of the pressure gradient and the Darcy-Weymouth nonlinear friction, to fast sound-wave controlled dynamics.

In addition to being uniquely positioned to interpolate directly, i.e. without any adaptation, between fast and slow regimes  the operator splitting method shows the following useful features.
\begin{itemize}
\item The explicit method is numerically accurate and stable,  e.g. in modeling transients over complex networks (in the networks with loops and in meshy networks).
\item The method matches performance of implicit methods, such as Kiuchi method \cite{Kiuchi1994} and the lump-element method \cite{ZCB2015},  in the slow regimes (native for the latter).
\item The method handles flawlessly and reliably effects of abrupt changes, generating multiple harmonics and fast sound-wave transients.
\item The method conserves total mass of gas (subject to the round-off error).
\end{itemize}
All the features of the split-step methods were discussed in the paper in details and also compared in multiple simulation tests against two other (implicit) methods -- the Kiuchi method~\cite{Kiuchi1994} and the lumped method~\cite{ZCB2015}.

We plan the following future studies, extending and generalizing results of this paper:
\begin{itemize}
\item Improving accuracy, e.g. making the split-step method of the fourth or even higher orders in the discretization step, is straightforward.

\item Solving cases with non-isothermal transients, that is accounting for dynamics of temperature, will require extensions from linear to curved/nonlinear characteristics -- implementable via additional interpolation sub-steps. Therefore, generally advantageous due to its simplicity explicit scheme could be developed for non-isothermal transients, which are typically simulated via implicit methods, see e.g. \cite{abbaspour10}.

\item This generalization beyond linear of the characteristic-propagation step  will also be useful for analysis of fast and intense transients, e.g. of emergency type, leading to a complicated multi-shock long-haul dynamics. In this regime, accounting for curved characteristics due to nonlinear advection term, dropped from the Euler equation in the basic model analyzed in this paper, will be needed. Notice, however,  that in this highly turbulent and under-damped regime an additional modeling input will be needed (possibly through experiments and/or 3d turbulent modeling) to validate a broad-range applicability of the Darcy-Weymouth term.
    
\item The choice of the operator split is by no means unique. In this paper we choose to work with the simplest split, which also have a clear physical meaning in the linear wave regime of fast low-intensity and weak dissipation transients. It will be of interest to choose another split, e.g. with one of the steps naturally adopted to a slow/adiabatic regime, of the type discussed in \cite{CBL2015}. More generally, analysis of how freedom in  splitting affects convergence, stability and complexity can lead, potentially, to new algorithms,  optimal for a specific regime (slow or fast) or a range of regimes.
\item Finally, one comprehensive open question we plan to address is if and how the split-step methodology may be useful for solving efficiently dynamic optimization and control problems of the type discussed in \cite{ZCB2015} and also generalizations accounting preventively for contingencies resulting potentially in a fast and devastating transients.
\end{itemize}

\section{Acknowledgments}
%DSA would like to thank Prof. Daniel~Apell\"o for fruitful discussion on numerical methods and application of method of characteristics to network nodes.

Authors would like to thank Daniel~Appel\"o, Scott Backhaus, Michele Benzi, Michael Herty, Vladimir Lebedev,  Sidhant Misra and Marc Vuffray for discussions and helpful suggestions.

This work was carried out at Los Alamos National Laboratory under the auspices of the
National Nuclear Security Administration of the U.S. Department of Energy under Contract No. DE-AC52-06NA25396, and was supported
by the Advanced Grid Modeling Research Program in the U.S. Department of Energy Office of Electricity Delivery and Energy Reliability, DTRA office of basic research and by Project GECO for the Advanced Research Project Agency-Energy of the U.S. Department of Energy under Award No. DE-AR0000673. The work of KAO was partially supported by NSh-9697.2016.2 during his visit to Landau Institute.

%\bibliographystyle{siam}
%\bibliography{gas}

\end{document}